\documentclass[
preprintnumbers,a4paper,superscriptaddress,aps,showpacs]{revtex4}
\usepackage{graphicx} 
\begin{document}
\preprint{EHU-FT/0104}
\def\la{\langle}
\def\ra{\rangle}
\def\om{\omega}
\def\Om{\Omega}
\def\vep{\varepsilon}
\def\wh{\widehat}
\newcommand{\beq}{\begin{equation}}
\newcommand{\eeq}{\end{equation}}
\newcommand{\beqa}{\begin{eqnarray}}
\newcommand{\eeqa}{\end{eqnarray}}
\newcommand{\intf}{\int_{-\infty}^\infty}
\newcommand{\into}{\int_0^\infty}
\def\toab{\wh{T}_{AB}}
\def\aap{\wh{a}_p}
\def\aaq{\wh{a}_{q}}
\def\aar{\wh{a}_{r}}
\def\acp{\wh{a}_p^\dag}
\def\acq{\wh{a}_{q}^\dag}
\def\acr{\wh{a}_{r}^\dag}
\def\xap{\wh{\psi}(X)}
\def\xcp{\wh{\psi}^\dag(X)}
\def\xcq{\wh{\psi}^\dag(y)}
\def\crossa{\wh{v}_\alpha(X)}
\def\crossc{\wh{v}_\alpha^\dag(X)}

%

%
\title{Quantum times of arrival for multiparticle states}
\author{A. D. Baute}
\affiliation{Fisika Teorikoaren Saila,
Euskal Herriko Unibertsitatea,
644 P.K., 48080 Bilbao, Spain}
\affiliation{Departamento de Qu\'\i mica-F\'\i sica,
Universidad del Pa\'\i s Vasco, Apdo. 644, Bilbao, Spain} 
\author{I. L. Egusquiza}
\affiliation{Fisika Teorikoaren Saila,
Euskal Herriko Unibertsitatea,
644 P.K., 48080 Bilbao, Spain}
\author{J. G. Muga}
\affiliation{Departamento de Qu\'\i mica-F\'\i sica,
Universidad del Pa\'\i s Vasco, Apdo. 644, Bilbao, Spain} 
\date{\today}
\begin{abstract}
Using the concept of crossing state and the formalism of second
quantization, we propose a prescription for computing the density of
arrivals of particles for multiparticle states, both in the free and
the interacting case.  The densities thus computed are positive,
covariant in time for time independent hamiltonians, normalized to the
total number of arrivals, and related to the flux.  We investigate the
behaviour of this prescriptions for bosons and fermions, finding boson
enhancement and fermion depletion of arrivals.
\end{abstract}
\maketitle
\section{Introduction}
A long standing issue in the theory and experiment of quantum
mechanics has been that of measuring and formalizing time observables.
In the last two decades a substantial body of work has been produced
clarifying theoretically and measuring experimentally quantities such
as dwell times \cite{Muga01}, tunneling times, or arrival times
\cite{ML00}.  In particular, many recent papers have challenged the
classical work of Allcock, who denied the possibility of defining a
quantum arrival-time concept \cite{Allcock69,Allcock69a,Allcock69b}.
In fact, these theoretical efforts and difficulties concerning arrival
times have been essentially decoupled from the daily practice of many
laboratories, where time-of-flight (TOF) methods are routinely
used. One reason for such a divorce is that, in most cases, a
classical analysis of the translational motion and the associated
arrival-time distribution is sufficient. It is now the case, however,
that the development of laser cooling techniques is bringing the
quantum nature of the atomic dynamics to the fore, thus approaching
the conditions for testing several proposed time-of-arrival (TOA)
theoretical distributions in a regime that differs from the classical
approximation.

Yet another difficulty for a comparison and further interaction
between experiment and theory is the absence, up to now, of a TOA
theory for multiparticle systems.  While the possibility of detecting
individual atoms with nanosecond time resolution in specific TOF
experiments is open \cite{RSBPNBWA01}, in the generic case the TOF
spectra are produced by clouds of many particles that may interact
with each other or/and with an external field.  The aim of this paper
xis to provide a quantum TOA theory which is applicable for the generic
(one dimensional) multiparticle case using the formalism of second
quantization (see for example \cite{Baym74}), together with the
crossing states introduced in \cite{BEMS00} and developed further in
\cite{BEM01b}. We shall also portray several numerical examples to
illustrate the phenomena of boson enhancement and fermion depletion of
common arrivals.

\section{Time of arrival of a single particle}

One of the major hindrances to the consideration of time observables
in the framework of standard quantum mechanics was Pauli's theorem,
which, simply put, states that no self-adjoint operator can exist that
has canonical commutation relations with a self-adjoint bounded or
semibounded hamiltonian, thus implying that the standard recipe
associating self-adjoint operators to observables cannot work for
time.

Nonetheless, Aharonov and Bohm considered the motion of free particles
as a clock to measure time, and introduced a time operator by
symmetrizing the classical expression for the time when a particle,
initially at the origin and with momentum $p$, passes point $x$, that
is, $t=mx/p$.  With a sign change this becomes the time of arrival at
the origin of a free particle that, at time $t=0$, is at position $x$
with momentum $p$ \cite{MSP98,MLP98}.  If the arrival occurs at $X$
rather than at the origin, then the corresponding ``Aharonov-Bohm
time-of-arrival operator'' takes the form
\beq
\toab(X)=\frac{m}{2}\left[(X-\wh{x})\frac{1}{\wh{p}}
+\frac{1}{\wh{p}}(X-\wh{x})\right]\,.
\eeq
For all practical purposes, this expression fulfills all the
properties one would expect of an operator associated with the
observable quantity time-of-arrival for free particles on the line
\cite{AB61} (for more details on this and the following topics, see
\cite{ML00}). It cannot be applied onto states with non vanishing 
zero momentum, which has been at times regarded as a drawback
\cite{GRT96,BR01}.  In fact, this ``difficulty'' is perfectly
physical, and mirrors the classical divergence of the time of arrival
when the particle's momentum tends to zero. It also explains how
Pauli's theorem can be circumvented: Aharonov and Bohm's time operator
$\toab$ is a maximally symmetric operator, therefore not self-adjoint.
Other steps had to be taken before maximally symmetric operators and
their concomitant POVMs (positive operator valued measures or
generalized non-orthogonal resolutions of the identity) were
understood physically, however.

Although the faith in Pauli's theorem could have been slightly shaken
by Aharonov and Bohm's proposal, the issue seemed to be settled after
the important series of papers of Allcock
\cite{Allcock69,Allcock69a,Allcock69b}, which apparently put to rest
all hope to obtain a sensible prescription for the quantum prediction
of times of arrival.  Even so, some adventurous souls kept on
searching for alternative formulations within quantum
mechanics. Kijowski in 1974 \cite{Kijowski74} put forward a procedure
to compute time-of-arrival probability densities for the free particle
case in a purely axiomatic way (see also similar later work by Werner
\cite{Werner86}).

With the advent of a better understanding of positive operator valued
measures (also known as generalized decompositions of the identity or
non-orthogonal measurements) \cite{SV81,Holevo82,Peres93,BGL95}, the
force of Pauli's argument was strongly diminished. In fact, it has
been possible to show the relation between Aharonov and Bohm's
time-of-arrival operator and Kijowski's distribution: they follow
naturally one from each other \cite{MPL99,EM00a}. $\toab$ is the first
operator moment of a POVM whose distribution function over a given
state is the corresponding Kijowski distribution. Kijowski's
distribution $\Pi(t)$ of times of arrival at $X$ for particles in a
state $|\psi(0)\ra\equiv |\psi(t=0)\rangle$, in the free particle
case, can be written as
\begin{equation}
\Pi(t,X)=\sum_{\alpha=\pm}|\langle\psi(0)|t,\alpha\rangle|^2\,,\label{pite}
\end{equation}
where $|t,\alpha\rangle$ are the two generalized eigenvectors of
$\toab$ with eigenvalue $t$, $\toab|t,\alpha\rangle=t|t,\alpha\rangle$
for $\alpha=\pm$. 
Note that the above mentioned domain ``difficulty'' of $\toab$ 
does not apply to the bilinear functional $\Pi(t)$, which may be defined for 
arbitrary physical states regardless of their behaviour at
$p=0$ \cite{MLP98}.  

The fact that $\toab$ is not a self-adjoint
operator is clearly identified from the non-orthogonality of the
complete basis $\left\{|t,\alpha\rangle\right\}_{t,\alpha=\pm}$.
The eigenvectors are related to each other by means of the 
relation  
\beq
|t,\alpha\ra=e^{i\wh{H}(t-t')/\hbar}|t',\alpha\ra
\eeq
which assures the invariance of the distribution with respect to time
traslations.  In particular,
\beq\label{talpha}
|t,\alpha\ra=e^{i\wh{H}t/\hbar}|v_{\alpha}\rangle,
\eeq
where we have used a special notation for the $t=0$ (generalized)
eigenvectors or ``crossing states'',
$|v_{\alpha}\rangle=|t=0,\alpha\ra$, where again $\alpha$ stands for
either $+$ or $-$ (we will not be denoting explicitly the point of
arrival $X$, which is part of the definition of these states, but it
is always implied).

In terms of these states we may rewrite Kijowski's distribution of
times of arrival at the point $X$ as
\begin{equation}
\Pi(t,X)=\sum_{\alpha=\pm}|\langle\psi(t)|v_{\alpha}\rangle|^2.
\label{pive}
\end{equation}
This also suggests a rewritting of the distribution in terms of an
operator for the density of arrivals at point $X$,
$\wh{\pi}(X)\equiv\sum_\alpha |v_{\alpha}\ra
\la v_{\alpha}|$,
\beq\label{pigen}
\Pi(t,X)=\la\psi(t)|\wh{\pi}(X)|\psi(t)\ra.
\eeq
Consider now the explicit form of the states $|v_{\alpha}\rangle$ in
momentum representation,
\begin{equation}
\langle p|v_{\alpha}\rangle=
\left(\frac{\alpha p}{h m}\right)^{1/2}\Theta(\alpha p)
e^{-ipX/\hbar}\,,\label{vmomen}
\end{equation}
where $\Theta(\cdot)$ is Heaviside's unit step function.  The correct
correspondence of $\Pi(t,X)$ with the classical case becomes now
evident. If the non commutativity of position and momentum operators
could be neglected, $\wh{\pi}$ would correspond to the sum of the
moduli of the fluxes that cross $X$ from both sides.  Also important
is the fact that in a classical setting the corresponding dynamical
variable provides the arrival distribution irrespective of the
dynamics and interaction potentials. In other words, Eq. (\ref{pigen})
generalizes the free motion case in a natural and simple way for
arbitrary interaction potentials, a task that could not be carried out
using the original axiomatic procedure of Kijowski or by quantizing
the classical time of arrival for each particular potential (the
expressions are not analytically known in general and pose formidable
ordering problems).

Underlying this rewriting of the time-of-arrival distribution a change
of emphasis is to be found: whereas in Eq. (\ref{pite}) the 
time-of-arrival distribution is obtained from the overlap of the
\emph{initial} wavefunction with the states associated with arrival at
the instant $t$, be it from the left ($\alpha=+1$) or the right
($\alpha=-1$), in Eq. (\ref{pive}) it is obtained as the overlap of
the \emph{evolved} wavefunction with the \emph{constant} states
$|v_{\alpha}\ra$ that
measure arrivals. 
The first point of
view is, in a way, predictive: given the initial state of the
particle, one can predict when the arrivals will occur.  
In the general case, with interacting potentials,
this view may also be adopted with $|t,\alpha\ra$ 
given by Eq.(\ref{talpha}), where the appropriate Hamiltonian 
is put in each case. 
{}From the second perspective, which could be
termed ``unconditional'', the arrival or otherwise of a particle at
$x=0$ is directly measured in physical space at every instant, using
local definitions that are in no way conditioned by the different
potentials in which the particles might be moving. 
This point of view, inspired by Wigner's formalization of the
time-energy uncertainty relation \cite{Wigner72}, was advocated in
\cite{BEMS00,BEM01b}, where the properties of the crossing states
$|v_{\alpha}\rangle$ were examined, and Eq. (\ref{pive}) was put
forward as an expression of density of arrivals also for the case of
interaction.  In \cite{BEM01b} we rewrote some other distributions
that had been proposed in the literature for time-of-arrival
distributions of particles in a potential (\cite{LJPU99}, later
superseded by \cite{LJPU00}; see also \cite{Leon00}) in terms of
crossing states, and showed that those defined in Eq. (\ref{vmomen})
were the only ones considered that led to classical correspondence
with the properties expected of such distributions.

Another particularly relevant aspect of the change of emphasis is that
it helps to understand that Eq. (\ref{pive}) need no longer be
normalized to unity. In which case $\Pi(t,X)$ is to be understood as a
density of arrivals of one particle: there might be a non zero
probability for the particle never arriving at $X$, or, if the
interacting potential were confining (such as the harmonic
oscillator), recurrences would appear corresponding to many different
arrivals. Notice that $\Pi(t,X)$ is a density of arrivals, not of
first arrivals only.

\section{Second quantization and time of arrival}

Even though TOF experiments with single atoms might be available in
not too distant a future, we need to understand better how to predict
time-of-arrival distributions for multiparticle systems. Most suited
for such a purpose is the formalism of second quantization.  One must
first realize that the distribution of arrivals is a property of the
same nature as the current density, or the kinetic enery, namely, it
is obtained as the sum of ``single particle'' contributions,
irrespective of the external or internal interactions affecting the
$N$-particle system.  This is a key observation to discard outright,
even for free motion, quantizations that would provide two-particle
terms.

Let $\aap$ and $\acp$ represent the annihilation and creation
operators that respectively eliminate and create a plane wave of
momentum $p$. Similarly, $\wh{\psi}(x)$ and $\wh{\psi}^\dagger(x)$ act
on the vacuum disposing of and creating a particle at point $x$. The
canonical commutation relations read
\begin{equation}
\left[\aap,\acq\right]_\pm=\delta(p-q)\,,\qquad{\rm and}\quad
\left[\wh{\psi}(x),\wh{\psi}^\dagger(y)\right]_\pm=\delta(x-y)\,,\label{ccr}
\end{equation}
where, as usual, $[,]_\pm$ stands for the commutator in the case of
bosons and for the anticommutator when fermions are involved. The
position operator is written as
\[\wh{x}=\int_{-\infty}^{+\infty}dx\,x\,
\wh{\psi}(x)\wh{\psi}^\dagger(x)\,,\]
and the inverse of the momentum operator as
\[\wh{p}^{-1}= \int_{-\infty}^{+\infty}dp\,\frac1p\, \acp\aap\,,\]
{}from which the following form for a generalization of the
time-of-arrival operator of Aharonov and Bohm might be inferred
($X=0$):
\begin{eqnarray}
\toab^{(1)}&=&-\frac{m}2\left(\wh{x}\wh{p}^{-1}+\wh{p}^{-1}\wh{x}\right)
=\nonumber\\
&=&-\frac{m}{4\pi\hbar}\intf dx\,dp\,dq\,dr\,
\frac{x e^{i(r-q)x/\hbar}}p \left[\left(\delta(p-q)+\delta(p-r)
\right)\acq\aar + 2 \acp\acq\aar\aap\right]\,.
\label{absecq}
\end{eqnarray}
In this expression one can recognize a one-particle component but also
a two-particle one. As pointed out above, this leads us to discard
this procedure, because of its unphysicality.

At any rate, $\toab$ is only valid for the free particle case, a
further limitation of this route.  The proper quantization procedure
for the multiparticle case starts, as noted above, by recognizing the
additive character of the time of arrival in terms of single particle
contributions. 

The basic  trick is  that for 
additive quantities taking the form of a sum of single particle
operators, 
\beq
\wh{G}=\wh{g}_1+\wh{g}_2+_....+\wh{g}_N\,,
\eeq
each of which has  matrix elements $g_{ji}=\la j|\wh{g}|i\ra$ in a complete
(single particle) basis, the multiparticle operator in second quantized
form is given by the simple expression
\beq
\wh{G}=\sum_{ij}g_{ji}a^\dagger_j a_i\,,
\eeq
where $a_i$ and $a^\dag_j$ are the i-th annihilation and j-th creation
operators. That is, they connect states $|i\rangle$ and $|j\rangle$
respectively with the vaccum state ($a_i|i\rangle=|0\rangle$ and
$|j\rangle=a^\dag_j|0\rangle$).

In the case of the arrival density operator we can directly apply
this procedure in momentum representation.  An even more compact
expression is obtained by using the crossing states, to generate
\emph{crossing operators}, both annihilation and creation. Consider
any (generalized) one particle state $|\varphi\rangle$. We can write
the annihilation and creation operators associated with the state as
\[
\wh{\varphi}=\intf dp\,\langle\varphi|p\rangle\aap\,;\qquad
\wh{\varphi}^\dag=\intf dp\,\langle p|\varphi\rangle\acp\,.
\]
On applying this procedure to the crossing states, we obtain the
crossing operators
\begin{eqnarray}
\crossa & = &  \int_{-\infty}^{+\infty}dp\,\langle v_{\alpha}|p\rangle\aap=
\int_{-\infty}^{+\infty}dp\,\left(\frac{\alpha p}{h m}\right)^{1/2}
\Theta(\alpha p) e^{ipX/\hbar}\aap\,;\label{crucea}\\
\crossc & = &  \int_{-\infty}^{+\infty}dp\,\langle p|v_{\alpha}\rangle\acp=
\int_{-\infty}^{+\infty}dp\,
\left(\frac{\alpha p}{h m}\right)^{1/2}\Theta(\alpha p)
e^{-ipX/\hbar}\acp\,.\label{crucec}
\end{eqnarray}
Let us now put together Eqs. (\ref{crucea}) and (\ref{crucec}) with
Eq. (\ref{pive}) to write the 
\emph{arrival density operator}
$\wh{\Pi}(X)$ for arrivals at $X$ in second quantized form, 
\begin{equation}
\wh{\Pi}(X)=\sum_{\alpha=\pm}\crossc\crossa=
\intf dp\,dq\,\frac{\sqrt{pq}}{h m}\Theta(pq)
e^{i(q-p)X/\hbar}\acp\aaq\,.\label{picero}
\end{equation} 
The left and right arrivals density operators $\wh{\Pi}_+(X)$ and
$\wh{\Pi}_-(X)$ are similarly defined as
\[
\wh{\Pi}_\alpha(X)=\crossc\crossa\,.
\]
We may also write from Eq. (\ref{picero}) 
the corresponding operator in Heisenberg 
picture, 
whose expectation value over the initial state will give us
the density of arrivals at point $X$ at instant $t$,
\beqa
\wh{\Pi}(t,X)&=&\wh{U}^\dag(t)\wh{\Pi}(X)\wh{U}(t)=
\sum_{\alpha=\pm} \wh{v}_\alpha^\dagger(X,t)  \wh{v}_\alpha(X,t)
\\
&=&
\intf dp\,dq\,\frac{\sqrt{pq}}{h m}\Theta(pq)
e^{i(q-p)X/\hbar}\acp(t)\aaq(t)\,,\label{pit}
\eeqa 
where $\acp(t)$ and $\aaq(t)$ are the time evolved creation and
annihilation operators, with evolution operator $\wh{U}(t)$,
i.e. $\acp(t)=\wh{U}^\dag(t)\acp \wh{U}(t)$
and similarly $\aap(t)=\wh{U}^\dag(t)\aap
\wh{U}(t)$.

The density of arrivals at instant $t$ at point $X$ for a generic
state $|\psi\rangle$ may thus be written as 
\[
\Pi(t,X;\psi)=\langle\psi (0)|\wh{\Pi}(t,X)|\psi (0)\rangle=
\la \psi(t)|\wh{\Pi}(X)|\psi(t)\ra\,.
\]
This expression agrees with Eq. (\ref{pive}) whenever $|\psi\rangle$
is a one particle state. Even though it is not immediately apparent
from expression (\ref{pit}) that we are obtaining positive
semidefinite distributions, this is indeed the case by construction:
$\wh{\Pi}(t,X)$ is a positive operator because it is a sum of two terms of
the form $\wh{A}^\dag\wh{A}$.

Furthermore, the one particle operator that is an extension of
Aharonov and Bohm's time-of-arrival operator at position $X$ for many
particles, even in the interacting case, is straightforwardly written
as
\[\wh{T}_X=\int_{-\infty}^{+\infty}dt\,t\,\wh{\Pi}(t,X)=
\intf dt\,dp\,dq\,\frac{\sqrt{pq}}{h m}\Theta(pq)
e^{i(q-p)X/\hbar}\,t\, \acp(t)\aaq(t)\,.
\]
By construction this is simply a one particle operator, which
coincides with $\toab$ over states whose content is just one free
particle.

If the evolution of the system is governed by the free particle
Hamiltonian it is easy to check that the integral over time of the
arrival-density operator $\wh{\Pi}^{\rm free}(t,X)$ sums to the total
particle number operator,
\[
\int_{-\infty}^{+\infty}dt\,\wh{\Pi}^{\rm free}(t,X)
=\int_{-\infty}^{+\infty}dp\,\acp\aap=\wh{N}\,.\]
This is no longer the case whenever the evolution operator is not the
free one; anyhow, we deduce from this expression that the arrival
density is normalized to the total number of arrivals. Notice
that in the interacting case the total number of arrivals need not
coincide with the total particle number, it may be smaller or bigger. 

A particularly important property of the arrival-density operator is
that the density of arrivals $\Pi(t,X;\psi)$ over any state is
covariant in time if the Hamiltonian is independent of time (as has
been assumed all along). Even though the properties of covariance,
positivity, and correct classical correspondence do not, by
themselves, completely fix the density of times of arrival, they are
minimal requirements, the lack of which would seriously impair any
proposal.

Even though in the presentation above we have restricted ourselves to
pure states, there is no problem in extending our proposal to mixed
states, as follows:
\[
\Pi(t,X;\wh{\rho})={\rm Tr}\left(\wh{\Pi}(t,X)\wh{\rho}(0)\right)=
{\rm Tr}\left(\wh{\Pi}(X)\wh{\rho}(t)\right)\,.
\]

For the sake of completeness, let us note down the flux operator for
many particles, in Schr\"odinger's picture,
\begin{eqnarray}
\wh{j}(X)&=&\frac{-i\hbar}{2m}
\left\{\xcp\partial_X\xap-\left[\partial_X\xcp\right]\xap\right\}=
\nonumber\\
&=&\frac1{2 h m}\intf dp\,dq\,e^{i(q-p)X/\hbar}(p+q)\acp\aaq\,.
\label{jotacerot}
\end{eqnarray}
or in Heisenberg's picture as 
\begin{equation}
\wh{j}(t,X)=\wh{U}^\dag(t)\wh{j}(X)\wh{U}(t)=\frac1{2 h m}
\intf dp\,dq\,e^{i(q-p)X/\hbar}(p+q)\acp(t)\aaq(t)\,,\label{jotat}
\end{equation}
(again assuming that the Hamiltonian is independent of time). Notice
that the flux, defined in this standard manner, is a one-particle
operator.

A straightforward comparison of Eqs. (\ref{pit}) and (\ref{jotat})
reveals the differences and similarities between $\wh{\Pi}$ and the
flux.  In the former a geometric mean of the momenta takes the place
of the arithmetic mean in the latter. Moreover, $\Pi$ counts the case
when $p$ and $q$ are both negative as a positive contribution to the
arrival density, whereas the same case counts as a negative flux
contribution in (\ref{jotat}).  This means that the quantity that
tends classically to the flux is $\Pi_+-\Pi_-$ rather than $\Pi$
itself.

\section{Free particles: boson enhancement and fermion depletion}

We have already made out several properties of the proposed
arrival-density operator, namely positivity, covariance, one-particle
status, classical limit, and normalization to total number of
arrivals. There is an obvious missing element yet, in that we have not
investigated so far whether the fermionic or bosonic character of the
particles involved is somehow reflected in the properties of the
distributions of times of arrival, as is to be expected.

In fact, this distinction between fermions and bosons is already
present in the proposed arrival-density distributions, as we will be
showing in this section. In order to portray this new property it is
enough to consider simply two-particle states, of generic form
\[
|\psi\rangle=\intf dp_1\,dp_2\,\psi(p_1,p_2)|p_1,p_2\rangle
=\frac1{\sqrt{2}}\intf
dp_1\,dp_2\,\psi(p_1,p_2)\acq\acp|0\rangle\,,\]
both for bosons and fermions,
where $|0\rangle$ is the vacuum state, and the normalization condition
reads
\[
\frac12\intf dp_1\,dp_2\,\left[\overline{\psi(p_1,p_2)}
\pm\overline{\psi(p_2,p_1)}\right]\psi(p_1,p_2)=1\,,
\]
where the upper sign corresponds to bosons and the lower one to fermions.

Consider $\psi(p_1,p_2)$ given as
\begin{equation}
\psi_\pm(p_1,p_2)=
\frac1{\sqrt{2\left(1\pm|\langle\chi_a|\chi_b\rangle|^2\right)}}
\left[\chi_a(p_1)\chi_b(p_2)\pm\chi_a(p_2)\chi_b(p_1)\right]\,,
\label{simanti}
\end{equation}
which fulfills the normalization requirement if $\chi_a$ and $\chi_b$
are normalized one particle wavefunctions. Quite obviously,
$\langle\chi_a|\chi_b\rangle$ stands for $\int
dp\,\overline{\chi_a(p)}\chi_b(p)$. In order to compare with the case
of distinguishable particles, we shall also be using
\begin{equation}
\psi_{\rm d}(p_1,p_2)=\chi_a(p_1)\chi_b(p_2)\,.\label{distin}
\end{equation}
%
%
%

Since the arrival-density operator $\wh{\Pi}(t,X)$ is a one-particle
operator, the density of arrivals over the state $|\psi_+\rangle$,
say, can be reorganized as
\[\langle\psi_+(0)|\wh{\Pi}(t,X)|\psi_+(0)\rangle=\frac1{N_+^2}
\sum_{i,j=a,b}\langle\chi_j|\chi_i\rangle\Pi_{ij}(t,X)\,,\]
where $N_+=\sqrt{2\left(1+|\langle\chi_a|\chi_b\rangle|^2\right)}$, and
\[\Pi_{ij}(t,X)=\sum_{\alpha=\pm}\langle\chi_i|\wh{v}_\alpha^\dag(X,t)
\wh{v}_\alpha(X,t)|\chi_j\rangle\,.
\]
Over the fermionic state $|\psi_-\rangle$ the cross terms carry a
negative sign in front.
The evolved crossing states are given by Eqs. (\ref{crucea}) and
(\ref{crucec}) on substituting $\aap$ and $\acp$ by $\aap(t)$ and
$\acp(t)$, respectively. 

On the other hand, the evaluation of the expectation value of the
evolved arrival density operator over the state $|\psi_{\rm
d}\rangle$, which computes the density of arrivals for two
distinguishable particles in such a state, produces just the two
diagonal terms, i.e.
\[\langle\psi_{\rm d}(0)|\wh{\Pi}(t,x)|\psi_{\rm d}(0)\rangle
=\Pi_{aa}(t,x)+\Pi_{bb}(t,x)\,.\]

It should be observed that these computations are general in that they
hold true for the case of interacting particles as well, as long as
the states have the form given above. These results indicate that
fermions and bosons (antisymmetric and symmetric states) present cross
terms in the density of arrivals completely analogous to those that in
spatial density signal the statistics of the particles.
In fact, the formalism of second quantization carries in itself the
fermionic or bosonic character of the particles concerned, through the
commutation relations.

As a consistency check one may  compute for the above states, 
$|\psi_\pm\ra$ and $|\psi_d\ra$, the corresponding reduced 
one particle density operators 
$\wh{\rho}^{(j)}$, $j=1,2$ ($\rho^{(1)}=\rho^{(2)}$ for 
$|\psi_\pm\ra$)
and note that in all three cases $\Pi(t,X)=\sum_{j} \Pi^{(j)}(t,X)$, 
where
\[\Pi^{(j)}(t,X)={\rm Tr}_j\left[\wh{\Pi}(t,X)
\wh{\rho}^{(j)}\right]\,.
\]
in agreement with the one-particle character of the arrival-time
distribution.

\begin{figure}
\includegraphics[height=8cm]{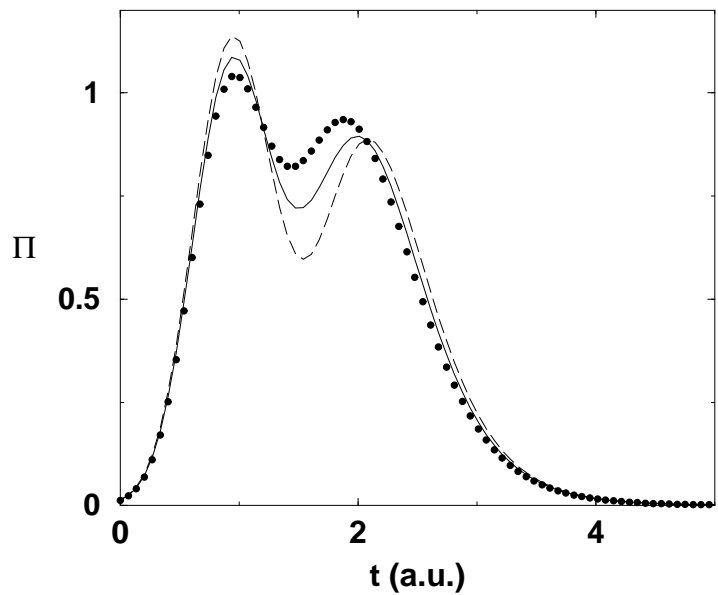}
\caption{\label{mapt2}
The solid line corresponds to distinguishable particles
($|\psi_d\rangle$), the dots show the density of arrivals for bosons
(symmetric state $|\psi_+\rangle$), and the dashed lines that of
fermions (antisymmetric state $|\psi_-\rangle$). The states are
defined by Eqs. \ref{simanti} and \ref{distin}, with $\chi_a$ and
$\chi_b$ gaussian states with minimum uncertainty at $t=0$, their
central positions being $\langle\chi_a|\wh{x}|\chi_a\rangle=-3.5$ and
$\langle\chi_b|\wh{x}|\chi_b\rangle=0$. In both cases, their central
positions in momentum space are at 3, and their spatial widths $\Delta
x=1$ (where $\Delta x$ is the square root of the spatial
variance). The point of observation is $X=3$, and the mass $m=1$. All
magnitudes are expressed in atomic units.}
\end{figure}

The difference between bosons, fermions and distinguishable particles
($|\psi_d\rangle$) is quite apparent in Fig. \ref{mapt2}. The one
particle states $\chi_a$ and $\chi_b$ are gaussians with a spatial
separation between them (in atomic units, $3.5$), while their width is
$1$ (a.u.).  Correspondingly, there are two main arrival times (maxima
of the arrival densities) for all three cases. Even so, the two sets
of principal arrivals for bosons (symmetric state) are much closer
together and much less differentiated than for distinguishable
particles ($|\psi_d\rangle$), which in turn present closer and less
differentiated maxima when compared to the fermionic (antisymmetric)
case.  It should be noticed that in this situation of free motion the
distributions are normalized to 2, as can be readily checked in this
numerical simulation.

\section{Interacting particles}
Consider  now a pair of interacting particles, be they
distinguishable, bosonic or fermionic, moving in otherwise free
space. The two-particle subspace of Fock space can be rewritten in
center of mass and relative coordinates, and we shall consider 
for simplicity factorized states of the form
\[
\psi(p_1,p_2)=\chi(P)\phi(p)\,,
\]
where $P=p_1+p_2$ is the center of mass momentum and $p=(p_1-p_2)/2$
the relative one. Under exchange of the particles $P$ is unchanged,
while $p$ flips sign. So in order to ensure that the state is bosonic
we are forced to use even functions $\phi_+(p)=\phi_+(-p)$, whereas
the fermionic case demands odd functions $\phi_-(p)=-\phi_-(-p)$. The
total mass is $2m$, while the reduced mass $\mu$ pertaining to the relative
system is $m/2$. The
normalization condition is translated into the requirement
that $\phi_\pm$ and $\chi(P)$ be normalized to unity.

In what follows we shall assume that the center of mass function is
gaussian with minimum uncertainty product at $t=0$. 
As to the internal states, they will be evolving in a
harmonic oscillator potential. We shall consider stationary and
coherent internal states.

\begin{figure}
\includegraphics[height=8cm]{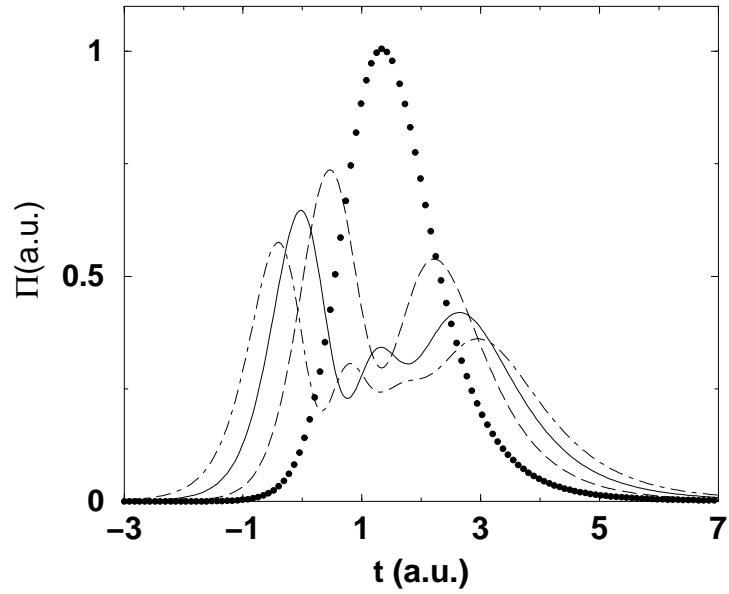}
\caption{\label{mapt4}
Arrival density of two identical particles in an internal stationary
state of the harmonic oscillator. The center of mass state is
gaussian, initially centered on $x=0$ and $p=4$ with width $\Delta
x=0.5$. Circles correspond to the (internal) ground state, dashes to
the first excited state, solid line to the second excited state,
dot-dash to the third one. The internal frequency is
$\omega=\sqrt{0.02}$ and the 
oscillation period is $T\approx 44.4$.
The point of crossing is $x=3$. All magnitudes
in atomic units.}
\end{figure}

\begin{figure}
\includegraphics[height=8cm]{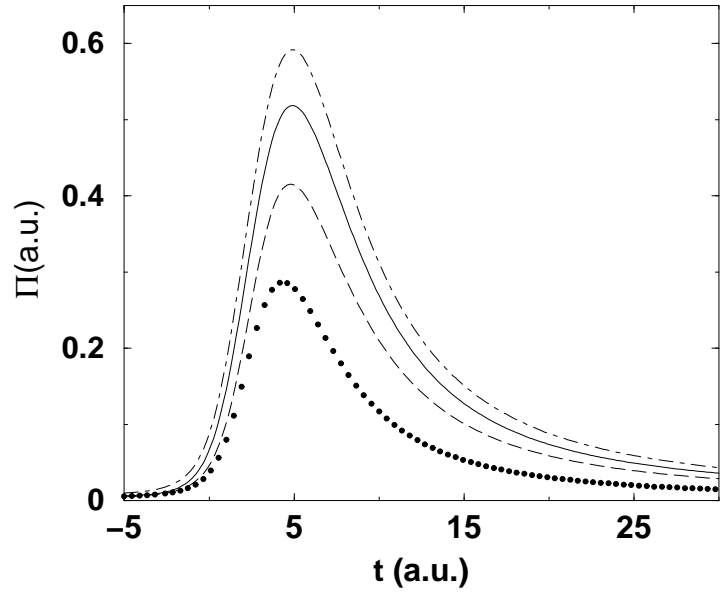}
\caption{\label{mapt5}
As in figure \ref{mapt4}, with the center of mass central momentum
changed to 1, $\Delta x=1$ and the internal frequency to 
$\omega=\sqrt{2}$.   
$T\approx 4.4$}
\end{figure}

Figs. \ref{mapt4} and
\ref{mapt5} represent two different sets of cases concerning
internal stationary states. The ground state and the even excited
states are symmetric (bosonic), whereas the odd numbered excited
states are antisymmetric (fermionic). The differences between
Figs. \ref{mapt4} and \ref{mapt5} are due to the different ratios
between internal energy and that of the center of mass motion. In
Fig. \ref{mapt4} the internal oscillations are much slower than the
center of mass motion, so the humps of the internal spatial
wavefunction appear, somewhat distorted for later times because of the
spreading, in the arrival density. However those humps are smoothed
over in Fig. \ref{mapt5} due to the much slower center of mass motion
relative to the internal motion.  Correspondingly, the integral of the
curves in Fig. \ref{mapt4} is very nearly 2, whereas there is a
significant increase of this number in Fig. \ref{mapt5} for the
excited states. The higher the excitation the broader the state is,
spatially, thus leading to more crossings.

We have also studied a case where the internal motion is time dependent. 
If it is fast enough with respect to the translational motion,
a peak structure corresponding to several  
oscillations may be observed.   
Let us consider, at time $t=0$,
symmetric and antisymmetric combinations of coherent
states, of the form
(remember that the coherent state $|z\rangle$ is
given by $\exp(-|z|^2/2)\sum_{n=0}^\infty(z^n/\sqrt{n!})|n\rangle$,
where $|n\rangle$ is the n-th excited state of the harmonic oscillator
hamiltonian)
\begin{equation}
|\phi_\pm\rangle=\frac1{\sqrt{2}}\left[|z\rangle\pm|
\bar{z}\rangle\right]\,,\label{coherent}
\end{equation}
where $\bar{z}$ is the complex conjugate of $z$.  We shall take in
particular $z=i$. In the relative motion space $|i\ra$ is a minimum
uncertainty product gaussian centered at the origin with average
momentum $(2\mu\omega\hbar)^{1/2}$ and spatial variance
$\hbar/(2\omega m)$. As time progresses it oscillates back and forth
along the relative motion coordinate with period $T=2\pi/\omega$.

\begin{figure}
\includegraphics[height=8cm]{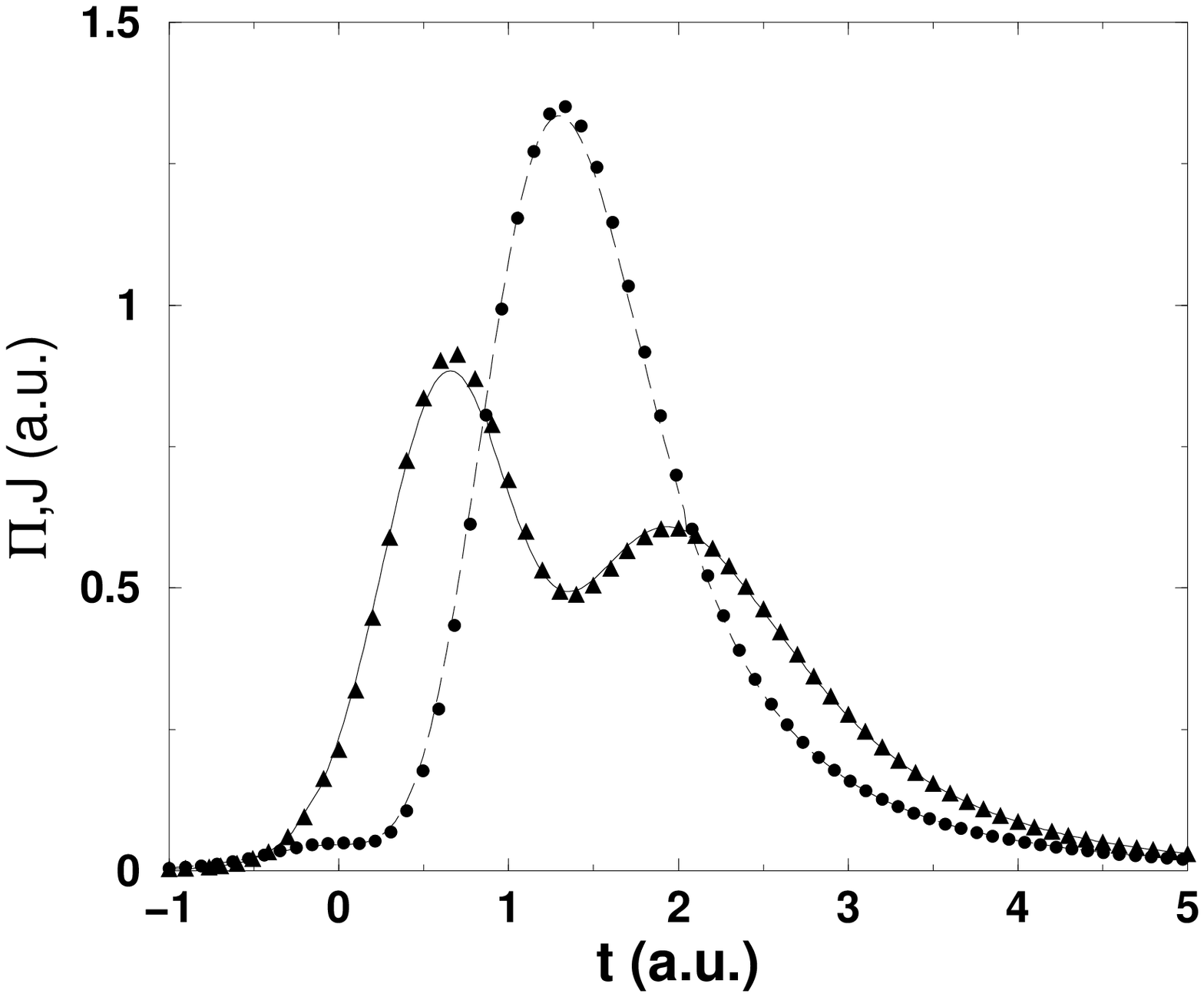}
\caption{\label{mapt7}
Arrivals density (solid line for fermions and dashed line for bosons)
and flux (triangles for fermions and circles for bosons) for the
internal states defined in Eq. (\ref{coherent}), with $z=i$ and internal
frequency $\omega=\sqrt{0.02}$. The initial center of mass state is a
gaussian with central position at $x=0$ and central momentum
$p=4$. The width is $\Delta x=0.5$. The point of arrival is $X=3$. All
magnitudes in atomic units.}
\end{figure}
In Fig. \ref{mapt7} the translational motion is faster than the
oscillations so we see just one peak for the symmetric case and two
maxima for the antisymmetric (fermionic) case. Since there is hardly
any component of negative momentum, there is no distinction between
flux and density of arrivals.

\begin{figure}
\includegraphics[height=8cm]{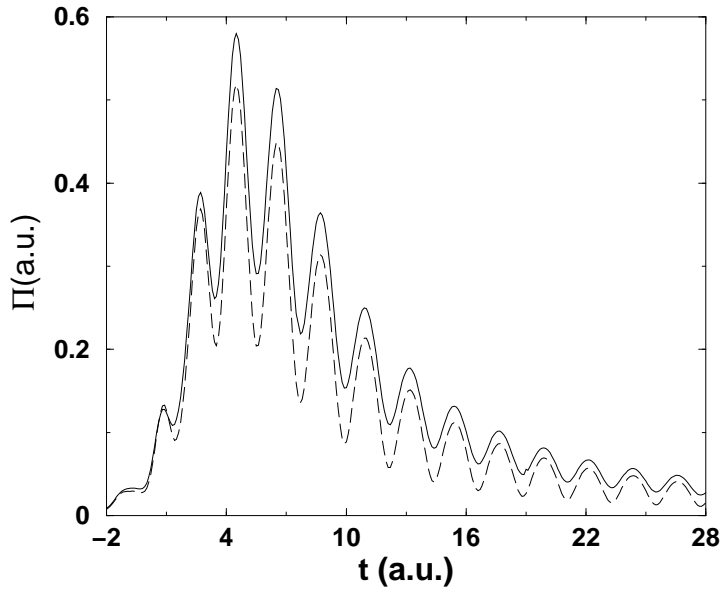}
\caption{\label{mapt8}
Arrivals density (solid line for fermions and dashed line for bosons)
for the internal states defined in Eq. (\ref{coherent}), with $z=i$ and
internal frequency $\omega=\sqrt{2}$.  The initial center of mass
state is a gaussian with central position at $x=0$, central momentum
$p=1$, and spatial width $\Delta x=1$. The point of arrival is
$X=3$. All magnitudes in atomic units.}
\end{figure}
In contrast to Fig. \ref{mapt7}, the internal potential is much
stronger in the situation depicted in Fig. \ref{mapt8}.  As opposed to
the case of stationary internal states, the tighter binding produces
here oscillations that can be clearly seen.

\section{Discussion}

In this work we have proposed a general method for computing densities
of arrivals (and related arrival density operators) for multiparticle
states, that fulfill a number of quite sensible demands: positivity,
covariance (if the evolution is homogeneous in time), related to a one
particle operator, normalized to the total number of particles in the
free case, related to the flux, and consistent with the classical 
arrival density. The analysis of the density of
arrivals (i.e., the one point function of the arrival density
operator) shows consistency with the results one would expect for
bosons, fermions, and distinguishable particles. Numerical
computations also show the behaviour expected, both for the free and
the interacting case, and reveal a number of physical effects,
hitherto unexplored.
The proposed distribution is also applicable to the case of 
external interaction potentials as shown already for the single particle  
states. 

The fact that the arrival-density operator is a one particle operator
implies that in fact no distinction is made for the one point function
(the density of arrivals) between the bosonic case and the symmetric
states of distinguishable particles. The full difference will be seen in
two-point and higher order functions, such as the arrival -
arrival  correlation function. In fact, on computing this two
point correlation function, one sees immediately that the one particle
component of the two point operator behaves in the natural way and can
be substracted from the arrival - arrival correlation
function to give the correlation function of pairs of arrivals,
$\langle\psi|:\wh{\Pi}(t)\wh{\Pi}(0):|\psi\rangle$, where $:\ :$
stands for normal ordering. The full analysis of this object we will
leave for future work.

In this paper we have not been overly concerned with domain problems
and the like. We know that the one particle time operator in the free
case is only maximally symmetric and cannot be made self-adjoint, and
we have no reason to expect that the interacting multiparticle case
will be simpler in this respect. Nonetheless, this is not particulary
relevant for our main interest, which lies in the computation of
densities of arrival. Notice furthermore that the fact that Aharonov
and Bohm's operator is not self-adjoint is really no hindrance to a
full quantum mechanical analysis of its associated densities.

There might be many other alternative prescriptions for times of
arrival. In the present state of knowledge, we do not feel able to
discard those outright. However, by pushing to the multiparticle case
the definitions used for one particle, this analysis becomes more
amenable to experimental test.

\begin{acknowledgments}
We acknowledge G. Hegerfeldt for useful comments.  This work is
supported by Ministerio de Educaci\'on y Cultura (AEN99-0315), The
University of the Basque Country (grant UPV 063.310-EB187/98), and the
Basque Government (PI-1999-28).  A. D.  Baute acknowledges an FPI
fellowship by Ministerio de Educaci\'on y Cultura.
\end{acknowledgments}


\begin{thebibliography}{26}
\expandafter\ifx\csname natexlab\endcsname\relax\def\natexlab#1{#1}\fi
\expandafter\ifx\csname bibnamefont\endcsname\relax
  \def\bibnamefont#1{#1}\fi
\expandafter\ifx\csname bibfnamefont\endcsname\relax
  \def\bibfnamefont#1{#1}\fi
\expandafter\ifx\csname citenamefont\endcsname\relax
  \def\citenamefont#1{#1}\fi
\expandafter\ifx\csname url\endcsname\relax
  \def\url#1{\texttt{#1}}\fi
\expandafter\ifx\csname urlprefix\endcsname\relax\def\urlprefix{URL }\fi
\providecommand{\bibinfo}[2]{#2}
\providecommand{\eprint}[2][]{\url{#2}}

\bibitem[{\citenamefont{Muga}(2001)}]{Muga01}
\bibinfo{author}{\bibfnamefont{J.~G.} \bibnamefont{Muga}}, in
  \emph{\bibinfo{booktitle}{Time in Quantum Mechanics}}, edited by
  \bibinfo{editor}{\bibfnamefont{J.~G.} \bibnamefont{Muga}},
  \bibinfo{editor}{\bibfnamefont{R.}~\bibnamefont{Sala-Mayato}},
  \bibnamefont{and} \bibinfo{editor}{\bibfnamefont{I.~L.}
  \bibnamefont{Egusquiza}} (\bibinfo{publisher}{Springer Verlag},
  \bibinfo{year}{2001}), \bibinfo{note}{to appear}, \eprint{quant-ph/0105081}.

\bibitem[{\citenamefont{Muga and Leavens}(2000)}]{ML00}
\bibinfo{author}{\bibfnamefont{J.~G.} \bibnamefont{Muga}} \bibnamefont{and}
  \bibinfo{author}{\bibfnamefont{C.~R.} \bibnamefont{Leavens}},
  \bibinfo{journal}{Phys. Rep.} \textbf{\bibinfo{volume}{338}},
  \bibinfo{pages}{353} (\bibinfo{year}{2000}).

\bibitem[{\citenamefont{Allcock}(1969{\natexlab{a}})}]{Allcock69}
\bibinfo{author}{\bibfnamefont{G.~R.} \bibnamefont{Allcock}},
  \bibinfo{journal}{Ann. Phys. (N.Y.)} \textbf{\bibinfo{volume}{53}},
  \bibinfo{pages}{253} (\bibinfo{year}{1969}{\natexlab{a}}).

\bibitem[{\citenamefont{Allcock}(1969{\natexlab{b}})}]{Allcock69a}
\bibinfo{author}{\bibfnamefont{G.~R.} \bibnamefont{Allcock}},
  \bibinfo{journal}{Ann. Phys. (N.Y.)} \textbf{\bibinfo{volume}{53}},
  \bibinfo{pages}{286} (\bibinfo{year}{1969}{\natexlab{b}}).

\bibitem[{\citenamefont{Allcock}(1969{\natexlab{c}})}]{Allcock69b}
\bibinfo{author}{\bibfnamefont{G.~R.} \bibnamefont{Allcock}},
  \bibinfo{journal}{Ann. Phys. (N.Y.)} \textbf{\bibinfo{volume}{53}},
  \bibinfo{pages}{311} (\bibinfo{year}{1969}{\natexlab{c}}).

\bibitem[{\citenamefont{Robert et~al.}(2001)\citenamefont{Robert, Sirjean,
  Browaeys, Poupard, Novak, Boiron, Westbrook, and Aspect}}]{RSBPNBWA01}
\bibinfo{author}{\bibfnamefont{A.}~\bibnamefont{Robert}},
  \bibinfo{author}{\bibfnamefont{O.}~\bibnamefont{Sirjean}},
  \bibinfo{author}{\bibfnamefont{A.}~\bibnamefont{Browaeys}},
  \bibinfo{author}{\bibfnamefont{J.}~\bibnamefont{Poupard}},
  \bibinfo{author}{\bibfnamefont{S.}~\bibnamefont{Novak}},
  \bibinfo{author}{\bibfnamefont{D.}~\bibnamefont{Boiron}},
  \bibinfo{author}{\bibfnamefont{C.~I.} \bibnamefont{Westbrook}},
  \bibnamefont{and} \bibinfo{author}{\bibfnamefont{A.}~\bibnamefont{Aspect}},
  \bibinfo{journal}{Science} \textbf{\bibinfo{volume}{292}},
  \bibinfo{pages}{461} (\bibinfo{year}{2001}).

\bibitem[{\citenamefont{Baym}(1974)}]{Baym74}
\bibinfo{author}{\bibfnamefont{G.}~\bibnamefont{Baym}},
  \emph{\bibinfo{title}{Lectures on Quantum Mechanics}} (\bibinfo{publisher}{W.
  A. Benjamin, Reading}, \bibinfo{address}{Massachusetts},
  \bibinfo{year}{1974}).

\bibitem[{\citenamefont{Baute et~al.}(2000)\citenamefont{Baute, Egusquiza,
  Muga, and {Sala Mayato}}}]{BEMS00}
\bibinfo{author}{\bibfnamefont{A.~D.} \bibnamefont{Baute}},
  \bibinfo{author}{\bibfnamefont{I.~L.} \bibnamefont{Egusquiza}},
  \bibinfo{author}{\bibfnamefont{J.~G.} \bibnamefont{Muga}}, \bibnamefont{and}
  \bibinfo{author}{\bibfnamefont{R.}~\bibnamefont{{Sala Mayato}}},
  \bibinfo{journal}{Phys. Rev. A} \textbf{\bibinfo{volume}{61}},
  \bibinfo{pages}{052111} (\bibinfo{year}{2000}), \eprint{quant-ph/9911088}.

\bibitem[{\citenamefont{Baute et~al.}(2001)\citenamefont{Baute, Egusquiza, and
  Muga}}]{BEM01b}
\bibinfo{author}{\bibfnamefont{A.~D.} \bibnamefont{Baute}},
  \bibinfo{author}{\bibfnamefont{I.~L.} \bibnamefont{Egusquiza}},
  \bibnamefont{and} \bibinfo{author}{\bibfnamefont{J.~G.} \bibnamefont{Muga}},
  \bibinfo{journal}{Phys. Rev. A} \textbf{\bibinfo{volume}{64}},
  \bibinfo{pages}{012501} (\bibinfo{year}{2001}), \eprint{quant-ph/0102005}.

\bibitem[{\citenamefont{Muga et~al.}(1998{\natexlab{a}})\citenamefont{Muga,
  {Sala Mayato}, and Palao}}]{MSP98}
\bibinfo{author}{\bibfnamefont{J.~G.} \bibnamefont{Muga}},
  \bibinfo{author}{\bibfnamefont{R.}~\bibnamefont{{Sala Mayato}}},
  \bibnamefont{and} \bibinfo{author}{\bibfnamefont{J.~P.} \bibnamefont{Palao}},
  \bibinfo{journal}{Superlattices Microstruct.} \textbf{\bibinfo{volume}{23}},
  \bibinfo{pages}{833} (\bibinfo{year}{1998}{\natexlab{a}}),
  \eprint{quant-ph/9801043}.

\bibitem[{\citenamefont{Muga et~al.}(1998{\natexlab{b}})\citenamefont{Muga,
  Leavens, and Palao}}]{MLP98}
\bibinfo{author}{\bibfnamefont{J.~G.} \bibnamefont{Muga}},
  \bibinfo{author}{\bibfnamefont{C.~R.} \bibnamefont{Leavens}},
  \bibnamefont{and} \bibinfo{author}{\bibfnamefont{J.~P.} \bibnamefont{Palao}},
  \bibinfo{journal}{Phys. Rev. A} \textbf{\bibinfo{volume}{58}},
  \bibinfo{pages}{4336} (\bibinfo{year}{1998}{\natexlab{b}}),
  \eprint{quant-ph/9807066}.

\bibitem[{\citenamefont{Aharonov and Bohm}(1961)}]{AB61}
\bibinfo{author}{\bibfnamefont{Y.}~\bibnamefont{Aharonov}} \bibnamefont{and}
  \bibinfo{author}{\bibfnamefont{D.}~\bibnamefont{Bohm}},
  \bibinfo{journal}{Phys. Rev.} \textbf{\bibinfo{volume}{122}},
  \bibinfo{pages}{1649} (\bibinfo{year}{1961}).

\bibitem[{\citenamefont{Grot et~al.}(1996)\citenamefont{Grot, Rovelli, and
  Tate}}]{GRT96}
\bibinfo{author}{\bibfnamefont{N.}~\bibnamefont{Grot}},
  \bibinfo{author}{\bibfnamefont{C.}~\bibnamefont{Rovelli}}, \bibnamefont{and}
  \bibinfo{author}{\bibfnamefont{R.~S.} \bibnamefont{Tate}},
  \bibinfo{journal}{Phys. Rev. A} \textbf{\bibinfo{volume}{54}},
  \bibinfo{pages}{4676} (\bibinfo{year}{1996}), \eprint{quant-ph/9603021}.

\bibitem[{\citenamefont{Briggs and Rost}(2001)}]{BR01}
\bibinfo{author}{\bibfnamefont{J.~S.} \bibnamefont{Briggs}} \bibnamefont{and}
  \bibinfo{author}{\bibfnamefont{J.~M.} \bibnamefont{Rost}},
  \bibinfo{journal}{Found. Phys.} \textbf{\bibinfo{volume}{31}},
  \bibinfo{pages}{693} (\bibinfo{year}{2001}).

\bibitem[{\citenamefont{Kijowski}(1974)}]{Kijowski74}
\bibinfo{author}{\bibfnamefont{J.}~\bibnamefont{Kijowski}},
  \bibinfo{journal}{Rept. Math. Phys.} \textbf{\bibinfo{volume}{6}},
  \bibinfo{pages}{361} (\bibinfo{year}{1974}).

\bibitem[{\citenamefont{Werner}(1986)}]{Werner86}
\bibinfo{author}{\bibfnamefont{R.}~\bibnamefont{Werner}}, \bibinfo{journal}{J.
  Math. Phys.} \textbf{\bibinfo{volume}{27}}, \bibinfo{pages}{793}
  (\bibinfo{year}{1986}).

\bibitem[{\citenamefont{Srinivas and Vijayalakshmi}(1981)}]{SV81}
\bibinfo{author}{\bibfnamefont{M.~D.} \bibnamefont{Srinivas}} \bibnamefont{and}
  \bibinfo{author}{\bibfnamefont{R.}~\bibnamefont{Vijayalakshmi}},
  \bibinfo{journal}{Pramana} \textbf{\bibinfo{volume}{16}},
  \bibinfo{pages}{173} (\bibinfo{year}{1981}).

\bibitem[{\citenamefont{Holevo}(1982)}]{Holevo82}
\bibinfo{author}{\bibfnamefont{A.~S.} \bibnamefont{Holevo}},
  \emph{\bibinfo{title}{Probabilistic and statistical aspects of quantum
  theory}} (\bibinfo{publisher}{North Holland, Amsterdam},
  \bibinfo{year}{1982}).

\bibitem[{\citenamefont{Peres}(1993)}]{Peres93}
\bibinfo{author}{\bibfnamefont{A.}~\bibnamefont{Peres}},
  \emph{\bibinfo{title}{Quantum Theory: Concepts and Methods}}
  (\bibinfo{publisher}{Kluwer}, \bibinfo{address}{Dordrecht},
  \bibinfo{year}{1993}).

\bibitem[{\citenamefont{Busch et~al.}(1995)\citenamefont{Busch, Grabowski, and
  Lahti}}]{BGL95}
\bibinfo{author}{\bibfnamefont{P.}~\bibnamefont{Busch}},
  \bibinfo{author}{\bibfnamefont{M.}~\bibnamefont{Grabowski}},
  \bibnamefont{and} \bibinfo{author}{\bibfnamefont{P.~J.} \bibnamefont{Lahti}},
  \emph{\bibinfo{title}{Operational quantum mechanics}}
  (\bibinfo{publisher}{Springer, Berlin}, \bibinfo{year}{1995}).

\bibitem[{\citenamefont{Muga et~al.}(1999)\citenamefont{Muga, Palao, and
  Leavens}}]{MPL99}
\bibinfo{author}{\bibfnamefont{J.~G.} \bibnamefont{Muga}},
  \bibinfo{author}{\bibfnamefont{J.~P.} \bibnamefont{Palao}}, \bibnamefont{and}
  \bibinfo{author}{\bibfnamefont{C.~R.} \bibnamefont{Leavens}},
  \bibinfo{journal}{Phys. Lett.} \textbf{\bibinfo{volume}{A253}},
  \bibinfo{pages}{21} (\bibinfo{year}{1999}), \eprint{quant-ph/9803087}.

\bibitem[{\citenamefont{Egusquiza and Muga}(2000)}]{EM00a}
\bibinfo{author}{\bibfnamefont{I.~L.} \bibnamefont{Egusquiza}}
  \bibnamefont{and} \bibinfo{author}{\bibfnamefont{J.~G.} \bibnamefont{Muga}},
  \bibinfo{journal}{Phys. Rev} \textbf{\bibinfo{volume}{A61}},
  \bibinfo{pages}{012104} (\bibinfo{year}{2000}), \bibinfo{note}{see also
  erratum, Phys. Rev. A {\bf 61} (2000) 059901(E)}, \eprint{quant-ph/9905023}.

\bibitem[{\citenamefont{Wigner}(1972)}]{Wigner72}
\bibinfo{author}{\bibfnamefont{E.~P.} \bibnamefont{Wigner}}, in
  \emph{\bibinfo{booktitle}{Aspects of quantum theory}}, edited by
  \bibinfo{editor}{\bibfnamefont{A.}~\bibnamefont{Salam}} \bibnamefont{and}
  \bibinfo{editor}{\bibfnamefont{E.~P.} \bibnamefont{Wigner}}
  (\bibinfo{publisher}{Cambridge University Press, London},
  \bibinfo{year}{1972}).

\bibitem[{\citenamefont{Le{\'{o}}n et~al.}(1999)\citenamefont{Le{\'{o}}n,
  Julve, Pitanga, and de~Urr{\'{\i}}es}}]{LJPU99}
\bibinfo{author}{\bibfnamefont{J.}~\bibnamefont{Le{\'{o}}n}},
  \bibinfo{author}{\bibfnamefont{J.}~\bibnamefont{Julve}},
  \bibinfo{author}{\bibfnamefont{P.}~\bibnamefont{Pitanga}}, \bibnamefont{and}
  \bibinfo{author}{\bibfnamefont{F.~J.} \bibnamefont{de~Urr{\'{\i}}es}}
  (\bibinfo{year}{1999}), \eprint{quant-ph/9903060}.

\bibitem[{\citenamefont{Le{\'{o}}n et~al.}(2000)\citenamefont{Le{\'{o}}n,
  Julve, Pitanga, and de~Urr{\'{\i}}es}}]{LJPU00}
\bibinfo{author}{\bibfnamefont{J.}~\bibnamefont{Le{\'{o}}n}},
  \bibinfo{author}{\bibfnamefont{J.}~\bibnamefont{Julve}},
  \bibinfo{author}{\bibfnamefont{P.}~\bibnamefont{Pitanga}}, \bibnamefont{and}
  \bibinfo{author}{\bibfnamefont{F.~J.} \bibnamefont{de~Urr{\'{\i}}es}},
  \bibinfo{journal}{Phys. Rev. A} \textbf{\bibinfo{volume}{61}},
  \bibinfo{pages}{062101} (\bibinfo{year}{2000}), \eprint{quant-ph/0002011}.

\bibitem[{\citenamefont{Le{\'{o}}n}(2000)}]{Leon00}
\bibinfo{author}{\bibfnamefont{J.}~\bibnamefont{Le{\'{o}}n}}
  (\bibinfo{year}{2000}), \eprint{quant-ph/0008025}.

\end{thebibliography}
\end{document}